# Hall Effect Mobility of Epitaxial Graphene Grown on Silicon Carbide

J.L. Tedesco, B.L. VanMil, R.L. Myers-Ward, J.M. McCrate, S.A. Kitt, P.M. Campbell[†], G.G. Jernigan[†], J.C. Culbertson[†], C.R. Eddy, Jr., and D.K. Gaskill

*Advanced Silicon Carbide Epitaxial Research Laboratory, U.S. Naval Research Laboratory, 4555 Overlook Avenue SW, Washington, D.C. 20375*
[†] *U.S. Naval Research Laboratory, 4555 Overlook Avenue SW, Washington, D.C. 20375*

## Abstract

Epitaxial graphene films were grown *in vacuo* by silicon sublimation from the (0001) and (000$\bar{1}$) faces of 4H- and 6H-SiC. Hall effect mobilities and sheet carrier densities of the films were measured at 300 K and 77 K and the data depended on the growth face. About 40% of the samples exhibited holes as the dominant carrier, independent of face. Generally, mobilities increased with decreasing carrier density, independent of carrier type and substrate polytype. The contributions of scattering mechanisms to the conductivities of the films are discussed. The results suggest that for near-intrinsic carrier densities at 300 K epitaxial graphene mobilities will be ~150,000 cm$^2$V$^{-1}$s$^{-1}$ on the (000$\bar{1}$) face and ~5,800 cm$^2$V$^{-1}$s$^{-1}$ on the (0001) face.

Graphene has been identified as a promising material for use in future electronic devices [1-3], including digital switches and radio-frequency transistors [4-5]. Thin films of graphene can be formed either through exfoliation of bulk graphite [6] or by epitaxial growth on SiC [3], with epitaxial graphene (EG) being more amenable to large-scale device processing [1-2]. Recently, EG has been patterned into devices [1,7-11] that have exhibited mobilities that are greater than or comparable to the mobilities of other high-mobility semiconductors including Si, Ge, GaN, and InAs [12-14]. However, most reports have neither cited the sheet carrier densities of the EG nor related those carrier densities to the mobilities in a systematic way. Such an approach would be of interest to determine if EG holds the promise of mobilities similar to recent reports of suspended exfoliated graphene at 300 K, e.g. ~200,000 $cm^2V^{-1}s^{-1}$ [15,16]. In this study, EG films were grown on semi-insulating SiC substrates and the Hall effect mobilities and sheet carrier densities were measured at both 300 K and 77 K to determine the relationship between the two properties. The conductivities of the EG films were also considered in order to begin to identify the scattering mechanisms affecting the mobilities. Understanding the relationship between mobility and carrier density as well as the scattering mechanisms limiting the film mobilities will be vital for designing the synthesis approach for future EG electronic devices.

Epitaxial graphene films were grown on 16 × 16 $mm^2$ samples sawed from both (0001) and (000$\bar{1}$) semi-insulating, on-axis 76.2 mm diameter 4H-SiC (Cree) and 6H-SiC (II-VI, Inc.) wafers. The samples were chemically cleaned *ex situ* [17] and *in situ* hydrogen etching and growth were conducted in a commercial hot-wall chemical vapor deposition (CVD: Epigress/Aixtron VP508) reactor. The *in vacuo* sublimation process [11] was used over a range of growth temperatures (1225 to 1700°C), vacuum pressures ($10^{-3}$ to $10^{-7}$ mbar), and growth times (10 to 120 minutes). Each synthesis experiment contained both faces and polytypes of SiC. Raman spectroscopy had been used previously to confirm the presence of graphene by observing the 2D peak at ~2700 cm$^{-1}$ [18] and a correlation was observed between a finite resistance of the epitaxial film and the observation of the 2D peak. Therefore, the presence of EG was normally confirmed by measuring the resistance across the sample.

Hall effect measurements were performed on 44 "large area" (16 × 16 mm$^2$) samples and 386 "small area" (10 × 10 μm$^2$) samples that were patterned on the large area samples using standard photolithography techniques. Hall effect mobilities, μ, and sheet carrier densities, n$_s$, of the large area samples were measured at 300 K and 77 K using copper pressure clips in a van der Pauw configuration, a magnetic field of ~2,060 G, and measurement currents of 1 to 100 μA. Small area samples were measured at 300 K only using similar conditions. To test reproducibility, samples were often re-measured several times and the data were found to be within 5% of the original values.

Plots of the Hall mobilities and carrier densities for large area EG and small area EG are shown in Fig. 1(a) and Fig. 1(b), respectively. The data displays that the mobilities increase with decreasing carrier density, independent of sample size, polytype, or dominant carrier type. Comparing Figs. 1(a) and 1(b), it is evident that the relationships between mobility and carrier density for large and small area samples are generally coincident. For comparison, mobility values and carrier densities from small area EG films synthesized *in vacuo* by others (0.5 × 6 μm$^2$ at 4 K [1], 0.6 × 3 mm$^2$ at 4 K [7], 0.1 × 1 mm$^2$ at 1.4 K [19], 1 × 6.5 μm$^2$ at 180 mK [20]) are plotted in Fig. 1(b). The data from these other studies lies within the scatter of the data from this study, suggesting that the functional relationship between increasing mobility and decreasing carrier density is an intrinsic property of EG synthesized via the sublimation technique. The mobilities and carrier densities for both large and small area samples were found to vary over a large range, but the carrier densities for Si-face EG was always lower than for C-face EG grown simultaneously. Furthermore, not all samples were n-type; about 40% of the samples exhibited holes as the dominant carrier, independent of growth face.

In Fig. 1(a), it appears that the lowest measured EG carrier densities are approaching the intrinsic carrier density, $n_i$, of graphene. The intrinsic carrier density can be determined using the equation $n_i = (\pi/6)(kT/\hbar v_F)^2$ [22], where k is Boltzmann's constant, T is the sample temperature, $\hbar$ is the reduced Planck constant, and $v_F$ is the Fermi velocity. The intrinsic carrier density has previously been determined to be $(8.5 \pm 0.5) \times 10^{10}$ cm$^{-2}$ at 300 K [22-23] and, based on the 300 K value, $n_i$ at 77 K was calculated to be $(5.5 \pm 0.5) \times 10^9$ cm$^{-2}$. It is evident that as $n_s$ approaches $n_i$, the projected 300 K mobility for C-face EG, increases to ~150,000 cm$^2$V$^{-1}$s$^{-1}$ and the projected 300 K mobility for Si-face EG, increases to ~5,800 cm$^2$V$^{-1}$s$^{-1}$. The projected C-face mobility is comparable to the reported intrinsic mobility of exfoliated graphene at 300 K due only to electron-phonon scattering, ~200,000 cm$^2$V$^{-1}$s$^{-1}$ [24] and the mobilities reported from suspended exfoliated graphene at 300 K, ~200,000 cm$^2$V$^{-1}$s$^{-1}$ [15-16]. Additionally, this suggests that the mobility of C-face EG at near-intrinsic carrier densities will be dominated by electron-phonon scattering.

The doping mechanism of EG is not understood at this time. Generally, the samples' dominant carrier type remained the same at 77 K as it had been at 300 K and the samples' mobilities generally increased as the temperature decreased. Also, the EG mobilities increase and the carrier densities decrease as the sample size decreases and whether samples were large or small, the highest hole mobilities are greater than the highest electron mobilities. The mobility increase at 77 K for the individual large area C-face samples was up to a factor of ~4 and the carrier densities decrease was by a factor of ~5.5 times relative to 300 K values. The changes were more dramatic for the Si-face samples, where mobilities increased by ~4 to 13.5 times, while the carrier densities decreased by ~6 to 28 times.

In an attempt to understand the observed differences in the temperature dependence of the mobilities, the conductivities, $\sigma$ ($\sigma = n_s e \mu$, where $e$ is the electron charge [26]), of the data found in Fig. 1 is shown in Fig. 2(a) for large area samples and Fig. 2(b) for small area samples. The C-face conductivities (at constant temperature) that are shown in Fig. 2 exhibit a sublinear dependence on carrier density for $n_s > 10^{13}$ cm$^{-2}$. The sublinear dependence is consistent with a scattering due to charged impurities, where $\sigma \sim n_s$ for $n_s \gg n_i$ [25,27], with other potential sublinear scattering mechanisms suppressing the conductivity. The behavior of the C-face conductivity plotted in Fig. 2 is in contrast to the behavior of the Si-face conductivity shown in Fig. 2. For the Si-face, both the large area and small area samples have nearly constant conductivity. This behavior is reminiscent of the scattering in exfoliated graphene due to ripples in the substrate, where $\sigma \sim n_s^{0.1}$ [25,28-29]. Therefore, the dependence of Si-face EG conductivity may result from inherently thinner films [3] and the well-known stepped morphology of EG graphene due to the underlying SiC substrate [11]. Furthermore, single or bilayer Si-face EG has an interfacial layer that differs significantly from the bonding between the C-face EG and the SiC substrate [30], which could give rise to interfacial point defects. Therefore, for $n_s \gg n_i$, it is also attractive to consider that further suppression of conductivity may be due to scattering from point defects in the interfacial layer, where $\sigma$ is expected to be constant with carrier density [25]. Interestingly, at 300 K, scattering from phonons should be the dominant scattering mechanism for epitaxial graphene on SiC [31], while scattering from charged impurities should dominate at lower temperatures [25]. However, the behavior of the large area C-face and Si-face conductivities shown in Fig. 2(a) remains the same at 77 K as at 300 K. This similar behavior suggests that the scattering mechanisms present at 300 K are not significantly affected by lowering the temperature only to 77 K; lowering the temperature further may be required to achieve a change in the conductivity behavior of the EG films.

Hall effect mobilities and sheet carrier densities were obtained for large area (16 × 16 mm$^2$) and small area (10 × 10 μm$^2$) EG films synthesized on SiC by the sublimation method. Unlike previous reports, sample with holes as dominant carriers were observed in EG films synthesized on both substrate faces. The mobilities were found to strongly depend on substrate face, and the mobilities were found to increase with decreasing carrier density on both faces, independent of carrier type or substrate polytype. Taken together, these results imply that it should be possible to produce graphene-based devices with mobilities approaching 150,000 cm$^2$V$^{-1}$s$^{-1}$ (5,800 cm$^2$V$^{-1}$s$^{-1}$) at 300 K when the carrier density of the C-face (Si-face) EG is reduced to intrinsic levels. Also, through comparison with results from other laboratories, the relationship between mobility and carrier density reported here may be an intrinsic property of EG synthesized via the sublimation technique.

**Acknowledgements**

This work was supported by the Office of Naval Research. JLT and BLV acknowledge support from the American Society for Engineering Education for Naval Research Laboratory Postdoctoral Fellowships. JMM and SAK acknowledge support from the Naval Research Enterprise Intern Program. The authors also acknowledge M.S. Fuhrer and Y. Liu for fruitful discussions.

**Figure Captions**

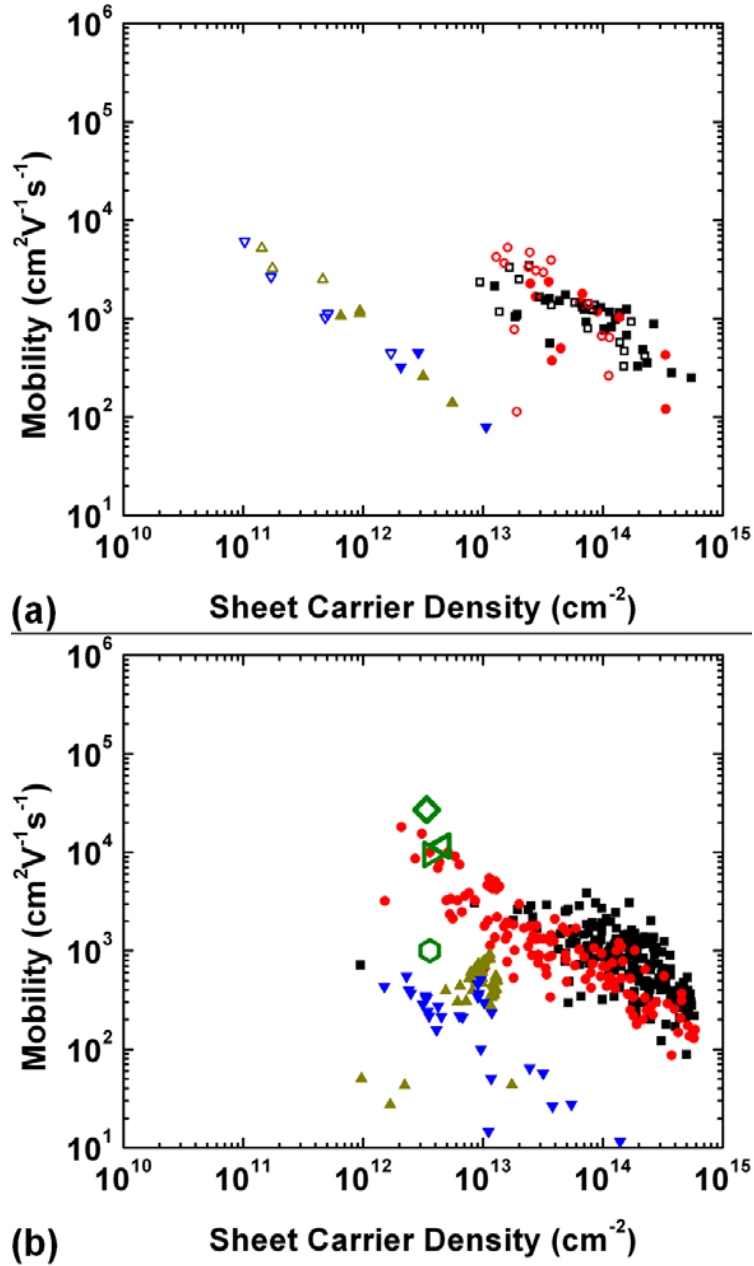

Fig. 1. (Color online) Epitaxial graphene Hall effect electron and hole mobilities and corresponding sheet carrier densities for (a) 16 × 16 mm$^2$ films at 300 K and 77 K and (b) 10 × 10 μm$^2$ samples patterned from the films in (a) and measured at 300 K only. Closed symbols are for 300 K sample data: C-face with electrons (■), C-face with holes (●), Si-face with electrons (▲), Si-face with holes (▼). Symbols for 77 K sample data: C-face with electrons at 77 K (□), C-face with holes at 77 K (○), Si-face with electrons at 77 K (△), and Si-face with holes at 77 K(▽). Data from other investigations correspond to other symbols: C-face with electrons at 180 mK (◇) [7], C-face with electrons at 1.4 K (◁) [19], C-face with electrons at 4.2 K (▷) [20], and a Si-face sample at 4 K with electrons [1] is an open hexagon.

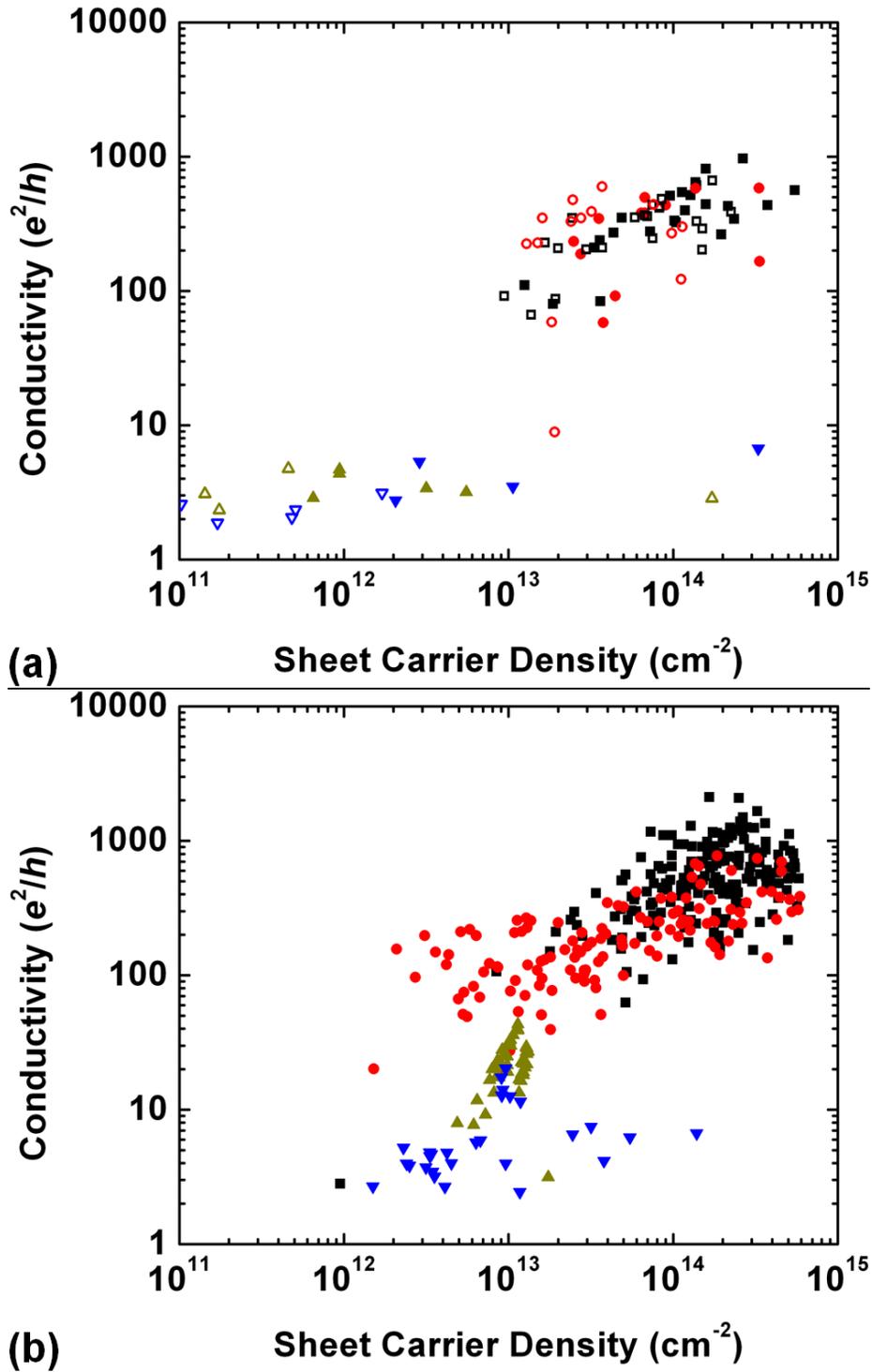

Fig. 2. (Color online) Calculated conductivities and corresponding sheet carrier densities for EG measured on (a) 16 × 16 mm$^2$ films at 300 K and 77 K and (b) 10 × 10 μm$^2$ samples patterned from the films in (a) and measured at 300 K only. Closed symbols are for 300 K sample data: C-face with electrons (■), C-face with holes (●), Si-face with electrons (▲), Si-face with holes (▼). Symbols for 77 K sample data: C-face with electrons at 77 K (□), C-face with holes at 77 K (○), Si-face with electrons at 77 K (△), and Si-face with holes at 77 K (▽).